\documentclass[12pt,preprint]{aastex}

\begin{document}

\title{Mixing and Transport of Short-Lived and Stable Isotopes and 
Refractory Grains in Protoplanetary Disks}

\author{Alan P.~Boss}
\affil{Department of Terrestrial Magnetism, Carnegie Institution for
Science, 5241 Broad Branch Road, NW, Washington, DC 20015-1305}
\email{boss@dtm.ciw.edu}

\begin{abstract}

 Analyses of primitive meteorites and cometary samples have
shown that the solar nebula must have experienced a phase of large-scale
outward transport of small refractory grains as well as homogenization
of initially spatially heterogeneous short-lived isotopes. The stable
oxygen isotopes, however, were able to remain spatially heterogenous
at the $\sim$ 6\% level. One promising mechanism for achieving these
disparate goals is the mixing and transport associated with a marginally 
gravitationally unstable (MGU) disk, a likely cause of FU Orionis events
in young low-mass stars. Several new sets of MGU models are presented 
that explore mixing and transport in disks with varied masses (0.016 to 
0.13 $M_\odot$) around stars with varied masses (0.1 to 1 $M_\odot$)
and varied initial $Q$ stability minima (1.8 to 3.1). The results show
that MGU disks are able to rapidly (within $\sim 10^4$ yr) achieve 
large-scale transport and homogenization of initially spatially 
heterogeneous distributions of disk grains or gas. In addition, the 
models show that while single-shot injection heterogeneity is reduced to 
a relatively low level ($\sim$ 1\%), as required for early solar system 
chronometry, continuous injection of the sort associated with the generation 
of stable oxygen isotope fractionations by UV photolysis leads to a 
sustained, relatively high level ($\sim$ 10\%) of heterogeneity, in 
agreement with the oxygen isotope data. These models support the suggestion
that the protosun may have experienced at least one FU Orionis-like
outburst, which produced several of the signatures left behind in primitive
chondrites and comets.

\end{abstract}


\section{Introduction}

 The short-lived radioisotope (SLRI) $^{26}$Al was alive during the formation
of the first refractory solids in the solar nebula, the Ca-, Al-rich inclusions
(CAIs) found in primitive chondritic meteorites. This means that at least some
of the solar system's SLRIs may have been injected into either the presolar 
cloud (e.g., Boss \& Keiser 2012; Boss 2012) or the solar nebula (Ouellette et 
al. 2007, 2010; Dauphas \& Chaussidon 2011) by a supernova or AGB star shock wave. 
In either case, injection occurred as a single event that was spatially 
heterogeneous, which would potentially reduce the usefulness of $^{26}$Al as a 
spatially homogeneous chronometer (Dauphas \& Chaussidon 2011) for precise 
studies of the earliest phases of planet formation (MacPherson et al. 2012; 
cf. Krot et al. 2012). Previous models (e.g., Boss 2011, 2012) have shown  how 
such initial spatial isotopic heterogeneity can be substantially reduced in a 
marginally gravitationally unstable (MGU) disk, as a result of the large-scale 
inward and outward transport and mixing of gas and particles small enough to move with 
the gas (e.g., Boss et al. 2012). Other elements and their isotopes suggest 
a similarly well-mixed solar nebula (e.g., Os: Walker 2012; Fe: Wang et al. 
2013). The stable oxygen isotopes, on the other hand, 
appear to have been spatially heterogeneous in the solar nebula during the early 
phases of planet formation; e.g., small refractory particles from Comet 81P/Wild 2
have normalized $^{17,18}$O/$^{16}$O ratios that span the entire solar system 
range of $\sim$ 6\% variations (Nakashima et al. 2012). The leading explanation
for generating these oxygen anomalies is UV photodissociation of CO molecules at 
the surface of the outer solar nebula (e.g., Podio et al. 2013), where 
self-shielding could lead to isotopic fractionation between gas-phase 
and solid-phase oxygen atoms (e.g., Lyons \& Young 2005; Krot et al. 2012). 
CO self-shielding on the irregular, corrugated outer surface of the disk 
would also lead to initial spatial heterogeneity, though the process would be 
continuous in time, rather than a single-shot event like a supernova shock wave.
Furthermore, the very existence of refractory particles in Comet 81P/Wild 2 
(Brownlee et al. 2006; Simon et al. 2008; Nakamura et al. 2008), which are
thought to have formed close to the protosun, implies that these small
particles experienced large-scale outward transport from the inner solar nebula
to the comet-forming regions of the outer solar nebula. MGU disks offer a
means to accomplish this early large-scale transport (e.g., Boss 2008, 2011;
Boss et al. 2012).

 Marginally gravitationally unstable disks are likely to be involved in
the FU Orionis outbursts experienced by young solar-type stars 
(e.g., Zhu et al. 2010b; Vorobyov \& Basu 2010; Martin et al. 2012). MGU 
disk models (e.g., Boss 2011) can easily lead to the high mass accretion rates 
($\sim 10^{-5} M_\odot$ yr$^{-1}$) needed to explain FU Orionis events.
FU Orionis outbursts are believed to last for about a hundred years and to 
occur periodically for all low mass protostars (Hartmann \& Kenyon 1996;
Miller et al. 2011). MGU models are also capable of offering an alternative 
mechanism (disk instability) for gas giant planet formation (e.g., Boss 2010; 
Meru \& Bate 2012; Basu \& Vorobyov 2012). However, the magnetorotational
instability (MRI) is likely to be involved in FU Orionis outbursts as well
(Zhu et al. 2009a), with MRI operating in the ionized innermost disk layers
as well as at the disk's surfaces. Zhu et al. (2009c, 2010a,b) have
constructed one- and two-dimensional (axisymmetric) models of a coupled
MGU-MRI mechanism, with MGU slowly leading to a build-up of mass in the
innermost disk, which then triggers a rapid MRI instability and an outburst.
Alternatively, MRI may operate in the outermost disk, partially ionized by
cosmic rays, leading to a build-up of mass in the dead zone at the intermediate
disk midplane, thus triggering a phase of MGU transport. Such a coupled
mechanism may be crucial for achieving outbursts in T Tauri disks,
where the disk masses are expected to be smaller than at earlier phases
of evolution.

 We present here several new sets of MGU disk models that examine the time evolution
of isotopic heterogeneity introduced in either the inner or outer solar nebula, 
by either a single-shot event or a continuous injection process, for a variety of disk
and central protostar masses, including protostars with M dwarf masses.
Low mass exoplanets are beginning to be discovered around an increasingly
larger fraction of M dwarfs (Bonfils et al. 2013; Dressing \& Charbonneau 2013; 
Kopparapu 2013), with a number of these being potentially habitable exoplanets,
elevating the importance of understanding mixing and transport processes in M dwarf
disks.

\section{Numerical Methods}

 The numerical models were computed with the same three dimensional,
gravitational hydrodynamics code that has been employed in previous MGU
disk models (e.g., Boss 2011). Complete details about the code and its
testing may be found in Boss \& Myhill (1992). Briefly, the code
performs second-order-accurate (in both space and time) hydrodynamics
on a spherical coordinate grid, including radiative transfer in the
diffusion approximation. A spherical harmonic ($Y_{lm}$)
expansion of the disk's density distribution is used to compute
the self-gravity of the disk, with terms up to and including 
$l = m = 32$. The radial grid contains 50 grid points for the 10 AU
disk models and 100 grid points for the 40 AU disk models. All
models have 256 azimuthal grid points, and effectively 45 theta
grid points, given the hemispherical symmetry of the grid. The theta
grid is compressed around the disk's midplane to provide enhanced
spatial resolution, while the azimuthal grid is uniformly spaced. The
Jeans length constraint is used to ensure adequate resolution. The
inner boundary absorbs infalling disk gas, which is added to the
central protostar, while the outer disk boundary absorbs the momentum
of outward-moving disk gas, while retaining the gas on the active grid.
The central protostar wobbles in such a manner as to preserve the
center of mass of the entire system.

 The time evolution of a color field is calculated (e.g., Boss 2011) 
in order to follow the mixing and transport of isotopes carried by the
disk gas or by small particles, which should move along with the disk gas. 
The equation for the evolution of the color field density $\rho_c$ 
(e.g., Boss 2011) is identical to the continuity equation for the disk gas 
density $\rho$

$${\partial \rho_c \over \partial t} + \nabla \cdot (\rho_c {\bf v}) = 0,$$

\noindent where ${\bf v}$ is the disk gas velocity and $t$ is the time. 
The total amount of color is conserved in the same way that the disk mass 
is conserved, as the hydrodynamic equations are solved in conservation
law form (e.g., Boss \& Myhill 1992).

\section{Initial Conditions}

 The initial disk density distributions are based on the approximation 
derived by Boss (1993) for a self-gravitating disk orbiting a star with 
mass $M_s$

$$ \rho(R,Z)^{\gamma-1} = \rho_o(R)^{\gamma-1} $$
$$ - \biggl( { \gamma - 1 \over \gamma } \biggr) \biggl[
\biggl( { 2 \pi G \sigma_o(R) \over K } \biggr) Z +
{ G M_s \over K } \biggl( { 1 \over R } - { 1 \over (R^2 + Z^2)^{1/2} }
\biggr ) \biggr], $$

\noindent where $R$ and $Z$ are cylindrical coordinates, $G$ is the gravitational
constant, $\rho_o(R)$ is the midplane density, $\sigma_o(R)$ is the surface density,
$K = 1.7 \times 10^{17}$ (cgs units) and $\gamma = 5/3$. 
The initial midplane density is

$$\rho_o(R) = \rho_{oi} \biggl( {R_{oi} \over R} \biggr)^{3/2}, $$

\noindent while the initial surface density is

$$\sigma_o(R) = \sigma_{oi} \biggl( {R_{oi} \over R} \biggr)^{1/2}. $$

\noindent The parameters $\rho_{oi}$ and $\sigma_{oi}$ and the
reference radius $R_{oi}$ are defined in Table 1 for the various
disk models explored in this paper. The total amount of mass in the
models does not change during the evolutions; the initial infalling
disk envelope accretes onto the disk, and no further mass is
added to the system across the outer disk boundary at $R_o$. The
outer disk surfaces are thus revealed to any potential source of
UV irradiation.

 For the 10 AU outer radius disks listed in Tables 2 and 3, the initial disk 
temperature profiles (Figures 1 and 2) are based on the Boss (1996) temperature 
profiles, with variations in the assumed outer disk temperature $T_o$, chosen 
in order to study the effect of varied minimum values of the $Q$
stability parameter. Values of $Q > 1.5$ indicate marginally gravitationally
unstable disks. The inner disks are all highly $Q$ stable, with $Q >> 1$.
For the 40 AU outer radius disks listed in Table 4, the initial 
disk temperatures are uniform at the specified outer disk temperature $T_o$,
leading to similar initial $Q$ values throughout the disks.
For all of the models, the temperature of the infalling envelope is 50 K.

 The initial color field is added to the surface of the initial disk in 
an azimuthal sector spanning either 45 degrees (10 AU outer radius disks)
or 90 degrees (40 AU outer radius disks) in a narrow ring of width 1 AU,
centered at the injection radii listed in the Tables. These models are intended to 
represent one-time, single-shot injections of isotopic heterogeneity, such as 
supernova-induced Rayleigh-Taylor fingers carrying live $^{26}$Al (e.g., 
Boss \& Keiser 2012). Table 4 lists both single-shot and continuous injection 
models, where in the latter case the color is added continuously to the same 
location on the disk surface throughout the evolution, crudely simulating 
ongoing photodissociation of CO (e.g., Lyons \& Young 2005) possibly leading 
to stable oxygen isotope fractionation between the gas and solid phases.
The color field in the latter case is intended to represent isotopically
distinct gas or small particles resulting from the UV photochemistry.
Note that in both the single-shot and continuous injection models, the
total amount of color added is arbitrary (e.g., the color field in the 
injection volume is simply set equal to 1), and is intended to be scaled
to whatever value is appropriate for the isotope(s) under consideration.
The color field is a massless, passive tracer that has no effect on the disk's
dynamics, so the total amount of color added is irrelevant for the disk's
subsequent evolution. The models seek to follow the deviations from
uniformity of the color field, not the absolute amounts of color added;
the evolution of the dispersion of the color field about its mean radial value,
divided by the mean radial value at each instant of time, is the goal of these 
models.

 Observations of the DG Tau disk by Podio et al. (2013) have shown that 
DG Tau itself irradiates its disk's outer layers from 10 AU to 90 AU with a
strong UV flux, sufficient for significant UV photolysis and the formation
of observable water vapor. Much higher levels of UV irradiation can occur
for protoplanetary disks that form in stellar clusters containing massive
stars (e.g., Walsh et al. 2013), an environment that has been suggested for
our own solar system (e.g., Dauphas \& Chaussidon 2011) in order to explain
the evidence for live SLRIs found in primitive meteorites.

 The fact that molecular hydrogen constitutes the great majority of a disk's
mass, yet cannot be directly detected, except at the star-disk boundary 
region, means that estimates of disk masses are uncertain at best
(e.g., Andrews \& Williams 2007), as they are typically based on an assumed 
ratio between the amount of mm-sized dust grains and the total disk mass.
Isella et al. (2009) estimated that low- and intermediate-mass pre-main-sequence
stars form with disk masses ranging from 0.05 to 0.4 $M_\odot$. DG Tau's 
disk has a mass estimated to be as high as 0.1 $M_\odot$ (Podio et al. 2013). 
Recently, the mass of the TW Hydra disk was revised upward to at least 0.05 $M_\odot$
(Bergin et al. 2013). These and other observations suggest that the MGU disk
masses assumed in these models may be achieved in some fraction of protoplanetary
disks, and perhaps in the solar nebula as well. In fact, Miller et al. (2011) 
detected a FU Orionis outburst in the classical T Tauri star LkH$\alpha$ 188-G4. 
Disk masses are typically thought to be $< 0.01 M_\odot$ for such stars. 
The fact that a FU Orionis event occurred in LkH$\alpha$ 188-G4 shows
that even the disks around Class II-type objects can experience instabilities 
leading to rapid mass accretion, e.g., MGU disk phases.

\section{Results}

 We present results for a variety of protostellar and protoplanetary disk masses,
varied initial minimum $Q$ stability parameters, and varied injection radii, for 
disks of two different sizes.

\subsection{10 AU Outer Radius Disks}

\subsubsection{G Dwarf Disks}

 Table 2 shows the initial conditions for the models with a $0.019 M_\odot$ 
disk in orbit around a $1.0 M_\odot$ protostar. The disks extend from 1 AU
to 10 AU, as in the models by Boss (2008, 2011). The main difference from
these previous models is that the disk mass ($0.019 M_\odot$) is considerably
lower than that of the previous models ($0.047 M_\odot$). As a result, the
initial minimum $Q$ values are considerably higher than in the previous
models, ranging from 2.2 to 3.1, compared to the previous range of 1.4 to
2.5. The present models are thus less gravitationally unstable initially
than the disks previously considered, with the goal being to learn whether
or not the previous results will change for higher values of $min\ Q_i$.
Figure 1 displays the initial midplane temperature profiles for these models.
Only the outermost regions of the disks are cool enough to be gravitationally
unstable, but the models show that this is sufficient to result in
qualitatively similar behavior for all of the Table 2 models.

 Figures 3 and 4 show the equatorial plane distribution of the color/gas
ratio ($\rho_c/\rho$) for model 1.0-2.6-9. This ratio is plotted, as
it is equivalent to the $^{26}$Al/$^{27}$Al and $^{17,18}$O/$^{16}$O 
ratios measured by cosmochemists, i.e., the abundance of an 
injected or photolysis product species, divided by that of a species 
that was prevalent in the pre-injection disk. Figure 3 shows that the initial
disk surface injection at 9 AU has resulted in the rapid transport of the 
color field downward to the disk's midplane, as well as inward to close to 
the inner disk boundary at 2 AU. The vigorous three dimensional motions
of a MGU disk are responsible for this large-scale transport in just 34 yr.
At this time, the color/gas ratio is still highly heterogeneous, but 
Figure 4 shows that only 146 yr later, the color/gas ratio has been
strongly homogenized throughout the entire disk midplane.

 Figure 5 shows the evolution of the dispersion of the ratio of the 
color density to the gas density for models 1.0-2.6-9 and 1.0-2.6-2 at two 
times. These models differ only in the injection radius, either 9 AU or 
2 AU. The dispersion plotted in Figure 5 is defined to be 
the square root of the sum of the squares of the color field divided
by the gas density, subtracted by the azimuthal average of this ratio at a
given orbital radius, divided by the square of the azimuthal average
at that radius, normalized by the number of azimuthal grid points,
and plotted as a function of radius in the disk midplane. Figure 5 shows 
that the isotopic dispersion is a strong function of orbital radius and time,  
with the dispersion initially being relatively large (i.e., at 180 yr, in spite
of the apparent homogeneity seen in Figure 4 at the same time) as a result of 
the isotopes traveling downward and radially inward and outward. However,
the dispersion decreases dramatically in the outer disks for both models 
by 777 yr to a value of $\sim$ 1\% to 2\%. In fact, the dispersion in
both models 1.0-2.6-9 and 1.0-2.6-2 evolves toward essentially the
same radial distribution by this time, showing that the exact location
of the injection location has little effect on the long term evolution
of the distribution: that is controlled solely by the evolution of the
underlying MGU disk, which is identical for these two models (i.e., the 
color fields are passive tracers, and have no effect on the disk's
evolution). Note that any small refractory grains present in the initial disk
will be carried along with the disk gas, so that some of the grains that 
start out at 2 AU will be transported to the outermost disk, in the same
manner that some of the gas is transported outward. Most of the gas and
dust, however, is accreted by the growing protostar.

 Figure 6 shows the results for three models with varied $min\ Q_i$, i.e.,
models 1.0-2.6-9, 1.0-2.9-9, and 1.0-3.1-9, all after 1370 yr. It can be
seen that in spite of the variation in the initial degree of instability,
the dispersions in the outermost disks all converge to similar values
of $\sim$ 1\% to 2\%. This suggests that MGU disk evolutions are not
particularly sensitive to the exact choice of the initial $Q$ profile,
a result that was also found by Boss (2011) for somewhat more massive disks.
As also found by Boss (2011), the dispersions in the innermost disks
are significantly higher ($\sim$ 10\% to 20\%) than in the outermost 
disks, a direct result of the stronger mixing associated with the cooler
outer disks, in spite of the longer orbital periods in the outer disks.

\subsubsection{M Dwarf Disks}

 Table 3 shows the initial conditions for the models with either $0.016 M_\odot$ 
disks around $0.1 M_\odot$ protostars, or $0.018 M_\odot$ disks around 
$0.5 M_\odot$ protostars. In either case, the disks extend from 1 AU to 10 AU.
These models are of interest for exploring how conditions might vary between
disks around G dwarfs and M dwarfs, with possible ramifications for the
habitability of any rocky planets that form (e.g., Raymond et al. 2007) around
M dwarfs. Figure 2 shows the initial midplane temperature profiles for these models.

 Figure 7 shows the time evolution of the dispersion for model 0.1-1.8-2, appropriate
for a late M dwarf protostar. As in all the models, it can be seen that the
initially highly heterogeneous disk becomes rapidly homogenized, in this case
by about 5000 yr. Note that this time scale is considerably longer than that
for G dwarf disk mixing and transport processes, as a result of the longer
Keplerian orbital periods for lower mass, M dwarf protostars.
As in the G dwarf protostar models (e.g., Figure 5), the inner 
disk dispersion is higher than in the outer disk, though in these models 
(with a lower initial $min\ Q = 1.8$) the inner disk dispersion drops to 
$\sim$ 5\% to 10\%, compared to $\sim$ 1\% to 2\% in the outer disk.
Figure 8 shows the same behavior for model 0.1-1.8-9, which differs from
the previous model shown in Figure 7 only in having the injection occur at
9 AU instead of 2 AU. As in the G dwarf disks, the dispersions for both
of these models evolve toward essentially identical radial distributions:
the underlying MGU disk evolution determines the outcome for the dispersions.
Similar results hold for the models with $0.5 M_\odot$ protostars, i.e.,
early M dwarf disks.
	
\subsection{40 AU Outer Radius Disks}

 We now turn to a consideration of the consequence of single-shot versus
continuous injection at the surface of much larger outer radius (40 AU) 
disks than have been considered to date for G dwarf stars; Boss (2007) 
considered disks extending from 4 AU to 20 AU in radius. Table 4 
shows the initial conditions for the models with a $0.13 M_\odot$ 
disk around a $1.0 M_\odot$ protostar, with the disks extending from 10 AU
to 40 AU. Because of the much larger inner and outer disk radii for
this set of models, these models can be calculated for times as long
as $\sim 3 \times 10^4$ yr (Table 4). Such times are still considerably
less than the typical ages ($\sim 10^6$ yr) of T Tauri stars, implying
that in order for MGU disks to occur at such late phases, a prior phase of
coupled MRI-MGU evolution might be required to make the present
results relevant.
 
 Figures 9 and 10 display the evolution of the dispersions for models
1.0-1.1-40-20 and 1.0-1.1-40-20c, differing only in that the former model
has single-shot injection while the latter model has continuous injection,
intended to simulate a disk with ongoing UV photolysis and fractionation
at the outer disk surface. For model 1.0-1.1-40-20, it can be seen
that the evolution is similar to that of the previous single-shot models:
a rapid drop in the dispersion, followed by homogenization to
$\sim$ 1\% to 2\% away from the inner disk boundary. The higher dispersions 
seen near the outer disk boundaries ($\sim$ 40 AU) are largely caused by the 
unphysical pile-up of considerable disk mass at 40 AU and should be discounted. 
However, for the continuous injection model shown in Figure 10, it can
be seen that the dispersions throughout the disk even after $\sim 10^4$ yr 
can be as high as $\sim$ 20\%, consistent with the much larger variation in 
stable oxygen isotope ratios, compared to SLRI ratios. In a calculation
with finer spatial grid resolution, as well as perhaps sub-grid mixing
processes, one might expect even stronger homogenization to occur, so
the dispersion levels obtained from the present models should be
considered to be upper bounds. The total amount of color added during
the continuous injection models is large, compared to single-shot
injection models: for model 1.0-1.4-40-20c, for example, after 200 yr, 
the total amount of color injected has increased by a factor of $\sim$ 90 
compared to the single-shot total, and by another factor of $\sim$ 60 
after 27000 yr.

 Figures 11 and 12 compare the results for continuous injection at either
20 AU or 30 AU, respectively, i.e., for models 1.0-1.1-40-20c and 
1.0-1.1-40-30c. In spite of the different injection radii, Figures 11
and 12 show that even at a relatively early phase (405 yr) of evolution,
the midplane color/gas ratios look somewhat similar; as before, the
MGU disk evolution is the same for both models, and that is the
primary determinant of the long term evolution. 

 Finally, similar results 
as those shown in Figures 9-12 were obtained for the other models 
listed in Table 4. These models show that the main factor in determining
the radial dispersion profile is whether the injection occurs in a
single-shot or continuously; in the latter case, the MGU disk does
its best to homogenize the color field, but the fact that spatial
heterogeneity is being continuously injected limits the degree
to which this heterogeneity can be reduced.

\section{Discussion}

 While dust grains in the interstellar medium are overwhelmingly 
amorphous, crystalline silicate grains have been found in a late M dwarf 
(SST-Lup3-1) disk at distances ranging from inside 3 AU to beyond 5 AU, 
in both the midplane and surface layers (Mer\'in et al. 2007). Such
crystalline silicate grains are likely to have been produced by 
thermal annealing in the hottest regions of the disk, well
inside of 1 AU (Sargent et al. 2009). Again, outward transport seems
to be required to explain the observations, and the results for
the models with a 0.1 $M_\odot$ protostar suggest that MGU phases
in low mass M dwarf disks may be responsible for these observations.
In fact, crystalline mass fractions in protoplanetary disks do not appear 
to correlate with stellar mass, luminosity, accretion rate, 
disk mass, or the disk to star ratio (Watson et al. 2009). These
results also appear to be consistent with the results of the present
models, which show that MGU disk phases are equally capable of relatively
rapid large-scale mixing and transport, regardless of the stellar or 
disk mass, or the exact value of the $Q$ stability parameter.

\section{Conclusions}

 These models have shown a rather robust result, namely that a phase
of marginal gravitational instability in disks and stars with a variety
of masses and disk temperatures can lead to relatively rapid inward and
outward transport of disk gas and small grains, as required to drive
the protostellar mass accretion associated with FU Orionis events, as
well as to explain the discovery of refractory grains in Comet 81P/Wild 2.
A MGU disk phase driving a FU Orionis outburst is astronomically quite 
likely to have occurred for our protosun, and cosmochemically convenient 
for explaining the relative homogeneity of $^{26}$Al/$^{27}$Al ratios 
derived from a supernova injection event, and the range of 
$^{17,18}$O/$^{16}$O ratios derived from sustained UV self-shielding 
at the surface of the outer solar nebula. Low-mass stars, from G dwarfs
to M dwarfs, may well experience a similar phase of MGU disk mixing
and transport. In this context, it is worthwhile to note that FU Orionis
itself, the prototype of the FU Orionis outburst phenomenon, has a
mass of only $\sim 0.3 M_\odot$ (Zhu et al. 2007, 2009b; Beck \& Aspin 2012), 
i.e., the mass of a M dwarf, suggesting that M dwarf protoplanetary disks 
may experience evolutions similar to that of the solar nebula, with
possible implications for the habitability of any resulting planetary
system (e.g., Raymond et al. 2007; Bonfils et al. 2013; Dressing \& 
Charbonneau 2013; Kopparapu 2013).

\acknowledgments

 I thank Jeff Cuzzi for his comments, the referee for a number of suggested
improvements, and Sandy Keiser, Michael Acierno, and Ben Pandit for their 
support of the cluster computing environment at DTM. This work was partially 
supported by the NASA Origins of Solar Systems Program (NNX09AF62G) 
and is contributed in part to the NASA Astrobiology Institute (NNA09DA81A). 
Some of the calculations were performed on the Carnegie Alpha Cluster, 
the purchase of which was partially supported by a NSF Major Research
Instrumentation grant (MRI-9976645).

\clearpage

\begin{deluxetable}{ccccccc}
\tablecaption{Density ($\rho_{oi}$) and surface density ($\sigma_{oi}$) parameters   
at a reference radius ($R_{oi}$) for varied disk masses ($M_d$) in orbit 
around varied mass protostars ($M_s$). $R_i$ and $R_o$ are the inner and outer
disk boundaries, respectively.  
\label{tbl-1}}
\tablehead{\colhead{\quad $M_s$ ($M_\odot$) \quad} &
\colhead{$M_d$ ($M_\odot$) \quad} &
\colhead{$\rho_{oi}$ \quad} &
\colhead{$\sigma_{oi}$ \quad} &
\colhead{$R_{oi}$ (AU) \quad} &
\colhead{$R_i$ (AU) \quad} &
\colhead{$R_o$ (AU) \quad} }
\startdata

1.0 &  0.019  &  $4.0 \times 10^{-10}$  & $3.2 \times 10^3$  &  1.0 &  1.0  &  10.0  \\

0.5 &  0.018  &  $3.1 \times 10^{-10}$  & $2.9 \times 10^3$  &  1.0 &  1.0  &  10.0  \\

0.1 &  0.016  &  $2.0 \times 10^{-10}$  & $2.5 \times 10^3$  &  1.0 &  1.0  &  10.0  \\

1.0 &  0.13   &  $1.0 \times 10^{-10}$  & $2.0 \times 10^3$  &  4.0 &  10.0 &  40.0  \\

\enddata
\end{deluxetable}

\clearpage

\begin{deluxetable}{ccccc}
\tablecaption{Initial conditions and final times ($t_f$) for models with a 10 AU 
outer radius, $0.019 M_\odot$ disk in orbit around a $1.0 M_\odot$ protostar. 
$T_o$ is the outer disk temperature, $min\ Q_i$ is the minimum value of
the initial $Q$ disk parameter, and $r_{inject}$ is the radius where
the color is injected at the disk's surface.
\label{tbl-2}}
\tablehead{\colhead{\quad \quad model \quad \quad} & 
\colhead{\quad \quad $T_o$ (K) \quad \quad} &
\colhead{\quad \quad $min\ Q_i$ \quad \quad} &
\colhead{\quad \quad $r_{inject}$ (AU) \quad \quad} & 
\colhead{\quad \quad $t_f$ (yr) \quad \quad} }
\startdata

1.0-2.2-2   &  15   &  2.2  &  2  &  2520  \\

1.0-2.2-9   &  15   &  2.2  &  9  &  2520  \\

1.0-2.6-2   &  20   &  2.6  &  2  &  2043  \\

1.0-2.6-9   &  20   &  2.6  &  9  &  2043  \\

1.0-2.9-2   &  25   &  2.9  &  2  &  1200  \\

1.0-2.9-9   &  25   &  2.9  &  9  &  1400  \\

1.0-3.1-2   &  30   &  3.1  &  2  &  1100  \\

1.0-3.1-9   &  30   &  3.1  &  9  &  1300  \\

\enddata
\end{deluxetable}

\clearpage

\begin{deluxetable}{ccccccc}
\tablecaption{Initial conditions and final times for models with 10 AU outer
radius disks ($M_d$) in orbit around lower mass protostars ($M_s$), as in Table 2. 
\label{tbl-3}}
\tablehead{\colhead{\quad model \quad} & 
\colhead{$M_s$ ($M_\odot$) \quad} &
\colhead{$M_d$ ($M_\odot$) \quad} &
\colhead{$T_o$ (K) \quad} &
\colhead{$min\ Q_i$ \quad} &
\colhead{$r_{inject}$ (AU) \quad} & 
\colhead{$t_f$ (yr) \quad} }
\startdata

0.1-1.8-2   &  0.1  &  0.016  &  40   &  1.8  &  2  &  8040  \\

0.1-1.8-9   &  0.1  &  0.016  &  40   &  1.8  &  9  &  8040  \\

0.5-2.4-2   &  0.5  &  0.018  &  25   &  2.4  &  2  &  4660  \\

0.5-2.4-9   &  0.5  &  0.018  &  25   &  2.4  &  9  &  4660  \\

\enddata
\end{deluxetable}

\clearpage

\begin{deluxetable}{cccccc}
\tablecaption{Initial conditions and final times for models with a 40 AU 
outer radius, $0.13 M_\odot$ disk in orbit around a $1.0 M_\odot$ protostar, as
in Table 2. 
\label{tbl-4}}
\tablehead{\colhead{\quad model \quad} & 
\colhead{\quad $T_o$ (K) \quad} &
\colhead{\quad $min\ Q_i$ \quad} &
\colhead{\quad $r_{inject}$ (AU) \quad} & 
\colhead{\quad injection mode \quad} & 
\colhead{\quad $t_f$ (yr) \quad} }
\startdata

1.0-1.1-40-20   &  30   &  1.1  &  20  &  single-shot &  25000  \\

1.0-1.1-40-30   &  30   &  1.1  &  30  &  single-shot &  24500  \\

1.0-1.4-40-20   &  50   &  1.4  &  20  &  single-shot &  24000  \\
 
1.0-1.4-40-30   &  50   &  1.4  &  30  &  single-shot &  15000  \\

1.0-1.1-40-20c  &  30   &  1.1  &  20  &  continuous  &  19500  \\

1.0-1.1-40-30c  &  30   &  1.1  &  30  &  continuous  &  19800  \\

1.0-1.4-40-20c  &  50   &  1.4  &  20  &  continuous  &  27000  \\

1.0-1.4-40-30c  &  50   &  1.4  &  30  &  continuous  &  27000  \\

\enddata
\end{deluxetable}

\clearpage

\begin{figure}
\vspace{-2.0in}
\plotone{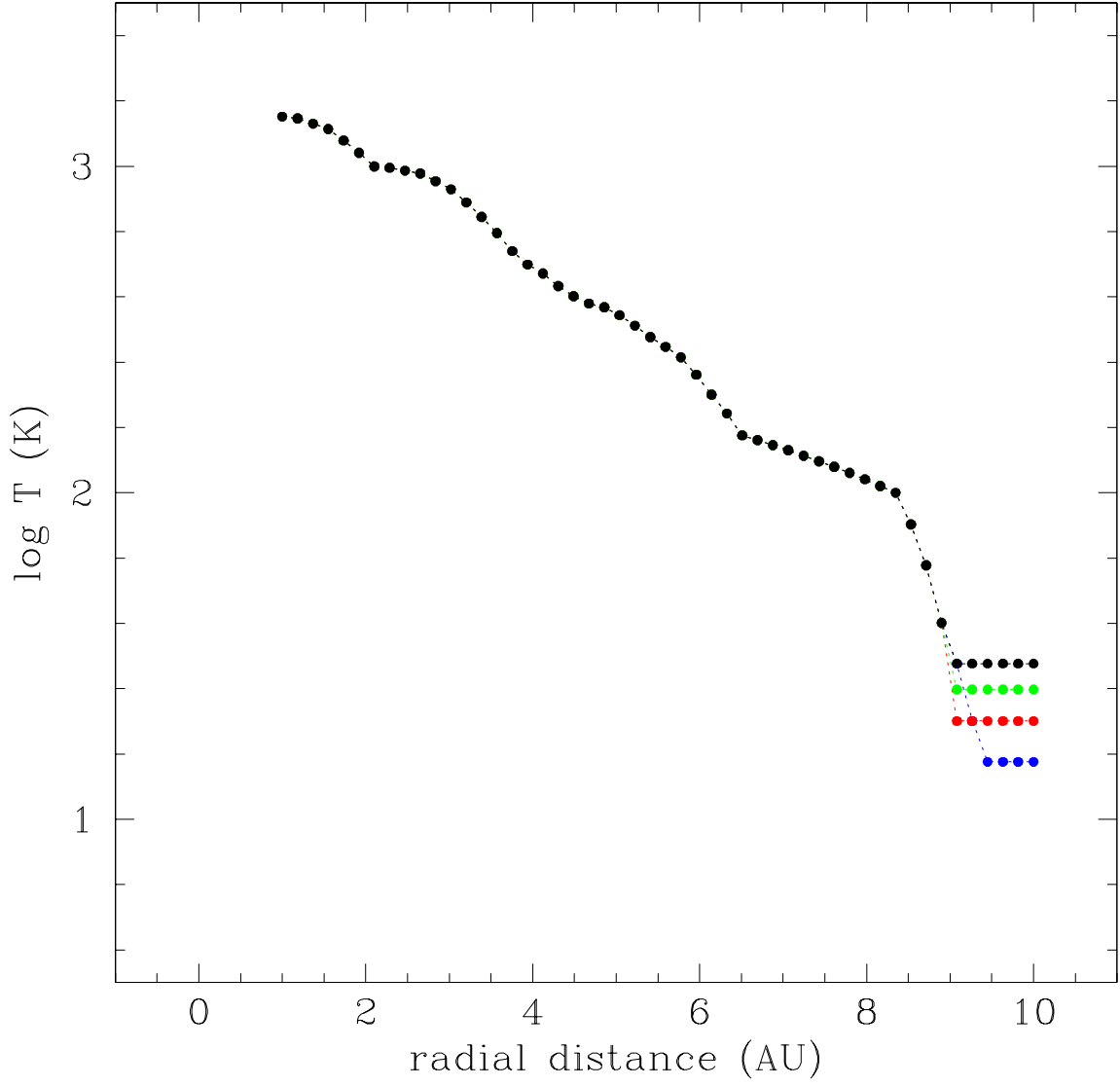}
\caption{Initial midplane temperature distributions (Boss 1996) for models
1.0-2.2-2 and 1.0-2.2-9 (blue), 1.0-2.6-2 and 1.0-2.6-9 (red),
1.0-2.9-2 and 1.0-2.9-9 (green), and 1.0-3.1-2 and 1.0-3.1-9 (black).}
\end{figure}

\clearpage

\begin{figure}
\vspace{-2.0in}
\plotone{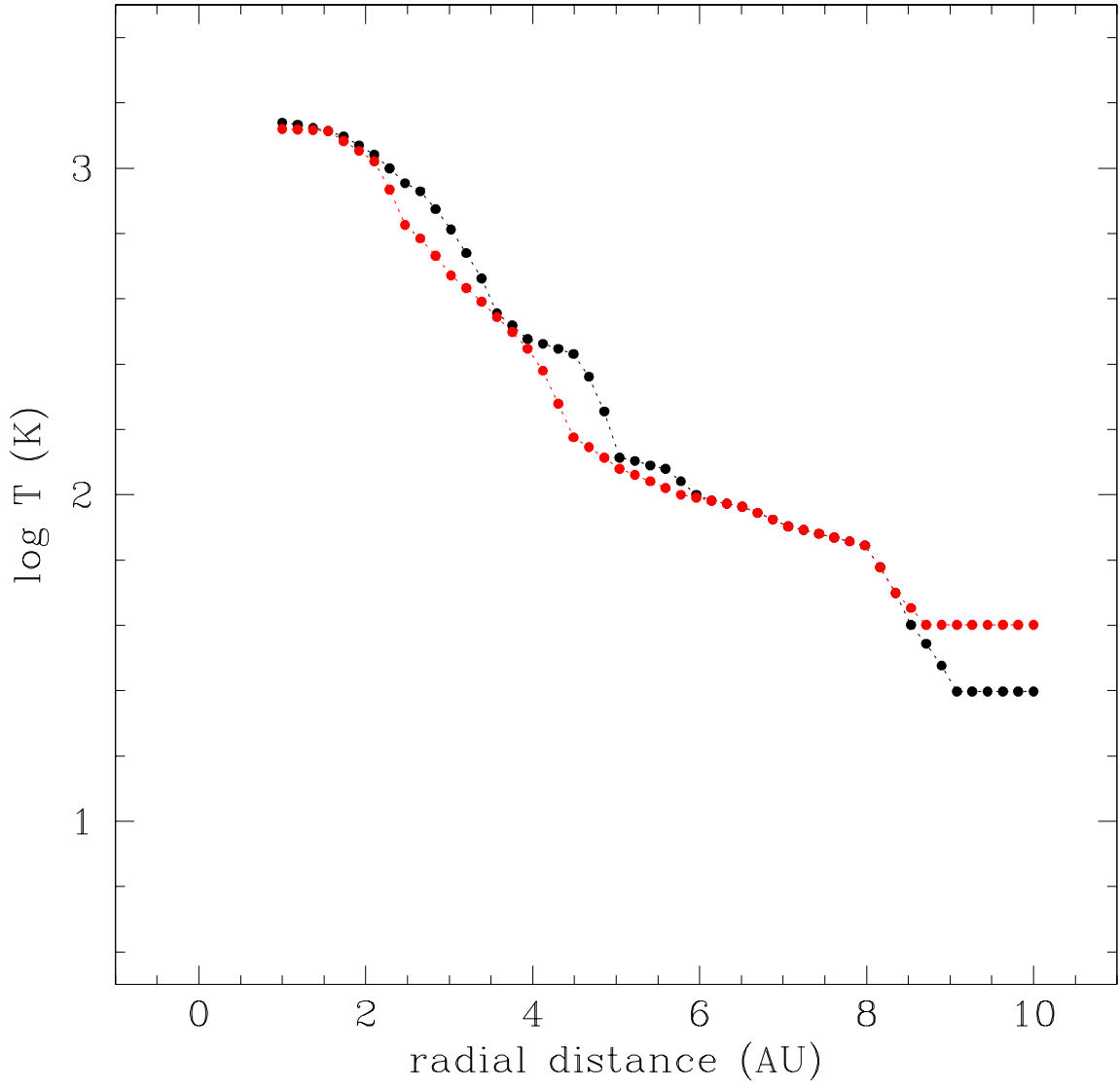}
\caption{Initial midplane temperature distributions (Boss 1996) for models
0.1-1.8-2 and 0.1-1.8-9 (red) and 0.5-2.4-2 and 0.5-2.4-9 (black).}
\end{figure}

\clearpage

\begin{figure}
\vspace{-1.0in}
\plotone{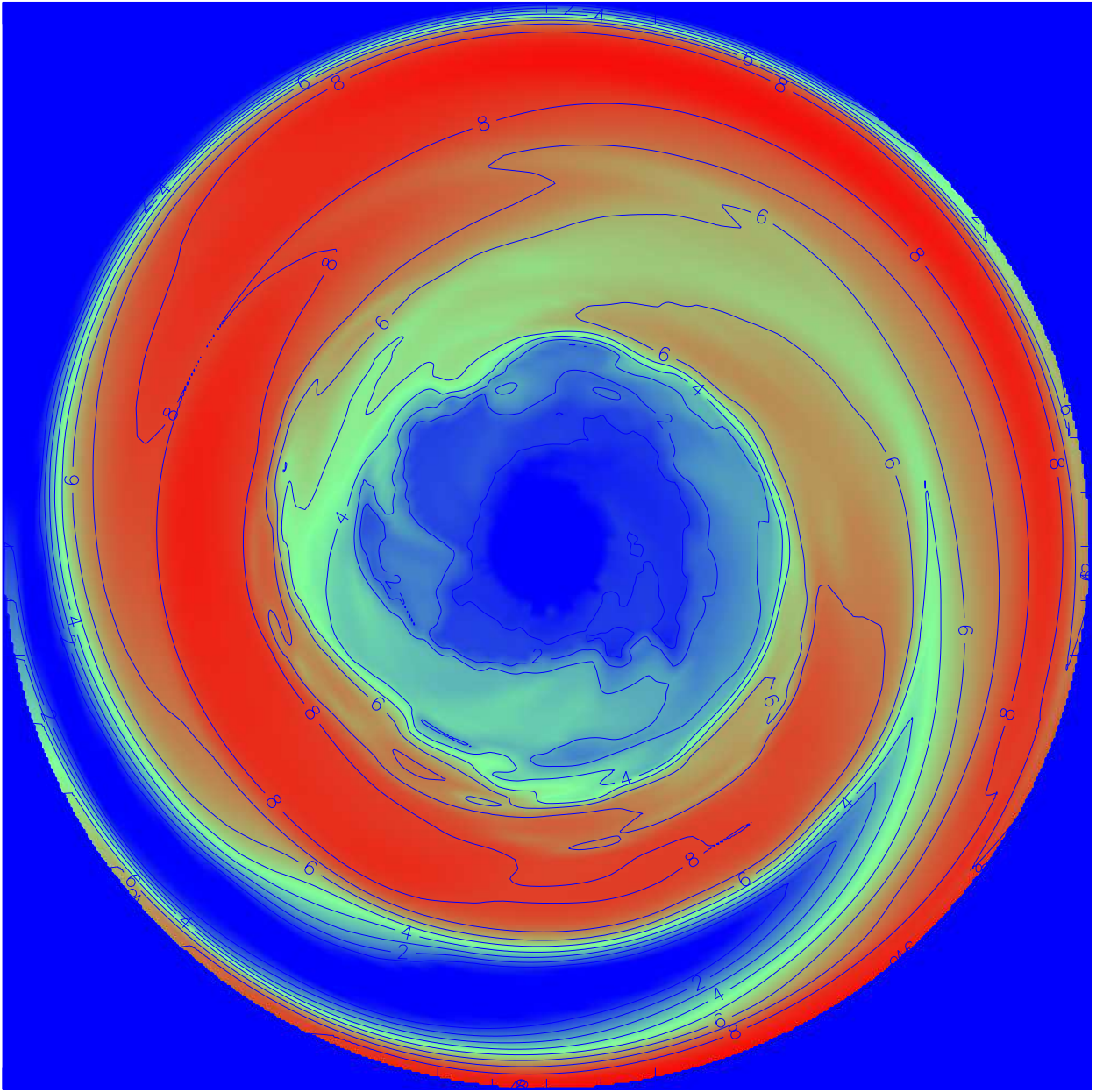}
\caption{Logarithm of the ratio of the color field to the gas density (log $\rho_c/\rho$)
in the disk's midplane (arbitrary units) after 34 yr for model 1.0-2.6-9, with single-shot
color injection at the initial disk's surface at 9 AU. Region
shown is 20 AU in diameter. A 1.0 $M_\odot$ protostar lies at the center of the MGU
disk. The color field has been transported inward to close to the inner boundary 
at 1 AU. The color to gas ratio is still highly heterogeneous.}
\end{figure}

\clearpage

\begin{figure}
\vspace{-1.0in}
\plotone{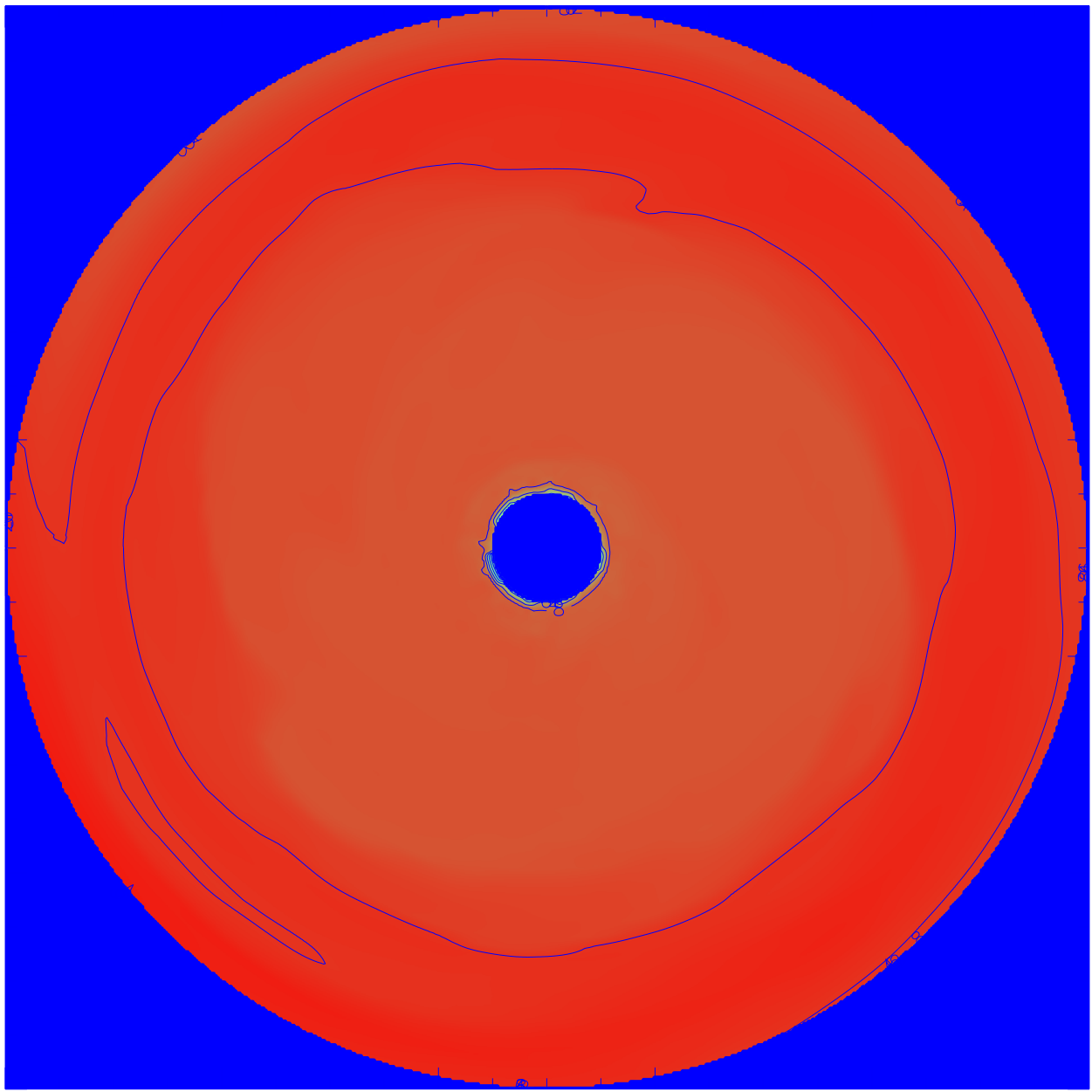}
\caption{Same as Figure 1, but after 180 yr. The color field has been transported
throughout the disk and the color to gas ratio is now relatively homogeneous.}
\end{figure}

\clearpage

\begin{figure}
\vspace{-2.0in}
\plotone{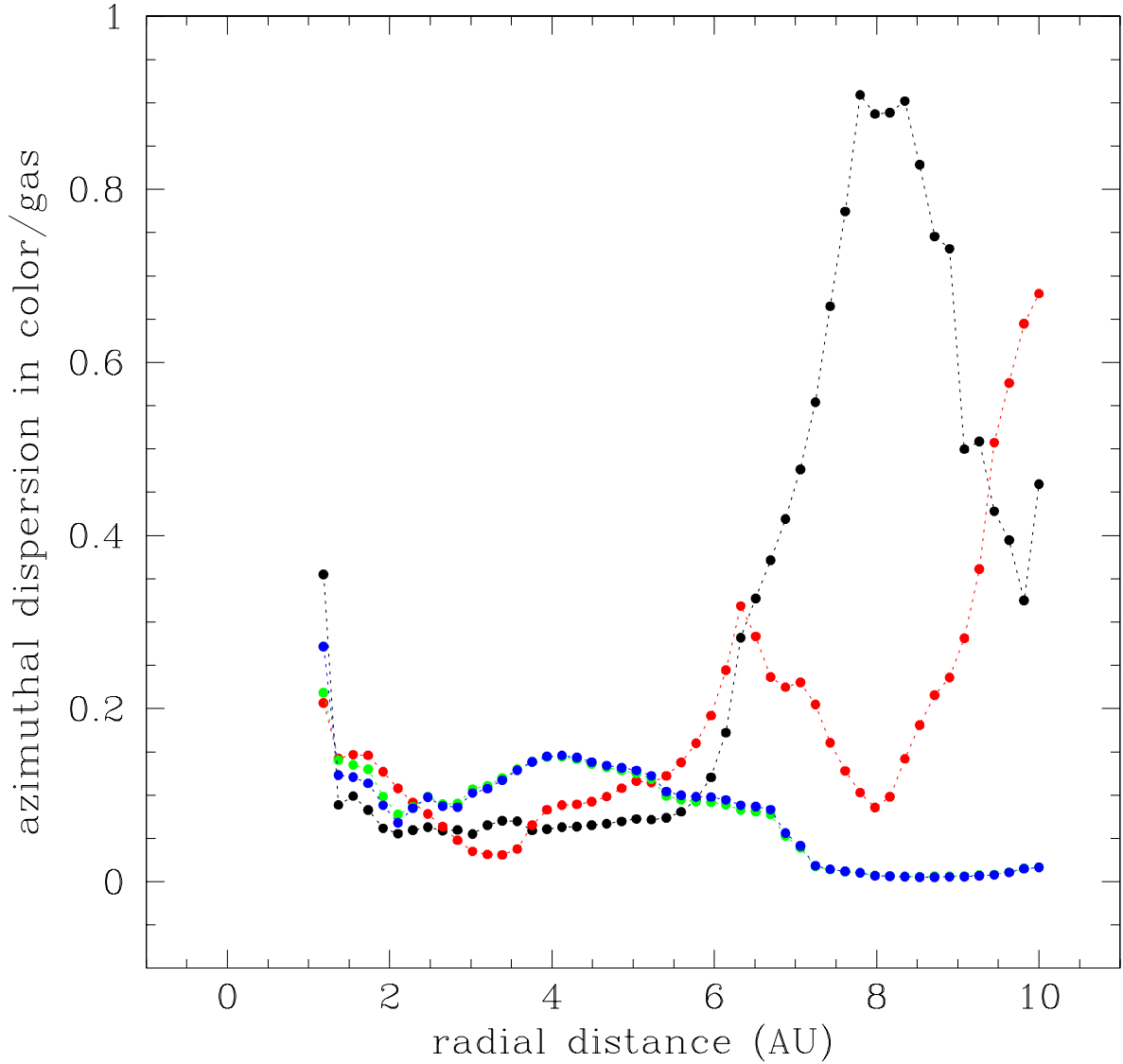}
\caption{Time evolution of the azimuthal dispersion of the color to gas ratio as a 
function of disk radius for models with single-shot surface injection at either 9 AU 
or 2 AU, respectively: model 1.0-2.6-9 (red at 180 yr, blue at 777 yr) and model 1.0-2.6-2 
(black at 180 yr, green at 777 yr). While initially quite different, the dispersions 
converge by 777 yr.}
\end{figure}

\clearpage

\begin{figure}
\vspace{-2.0in}
\plotone{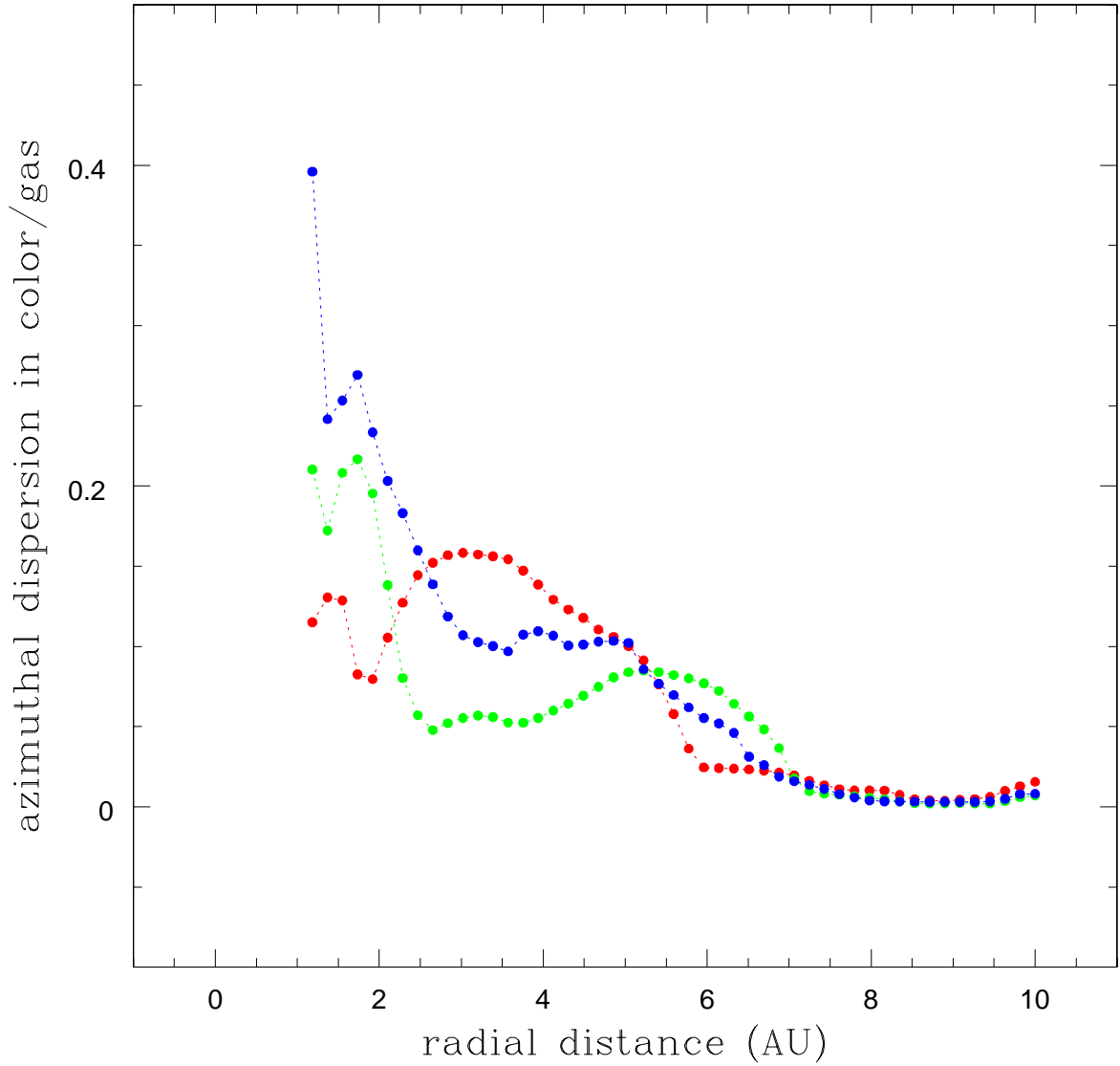}
\caption{Comparisons of the azimuthal dispersions after 1370 yr of the color to gas 
ratio as a function of disk radius, for models with single-shot surface injection at 9 AU,
but varied initial minimum $Q$ values of 2.6, 2.9, and 3.1: models 1.0-2.6-9 (red), 
1.0-2.9-9 (green), and 1.0-3.1-9 (blue), respectively. The outer disks have become
relatively homogenized.}
\end{figure}

\clearpage

\begin{figure}
\vspace{-2.0in}
\plotone{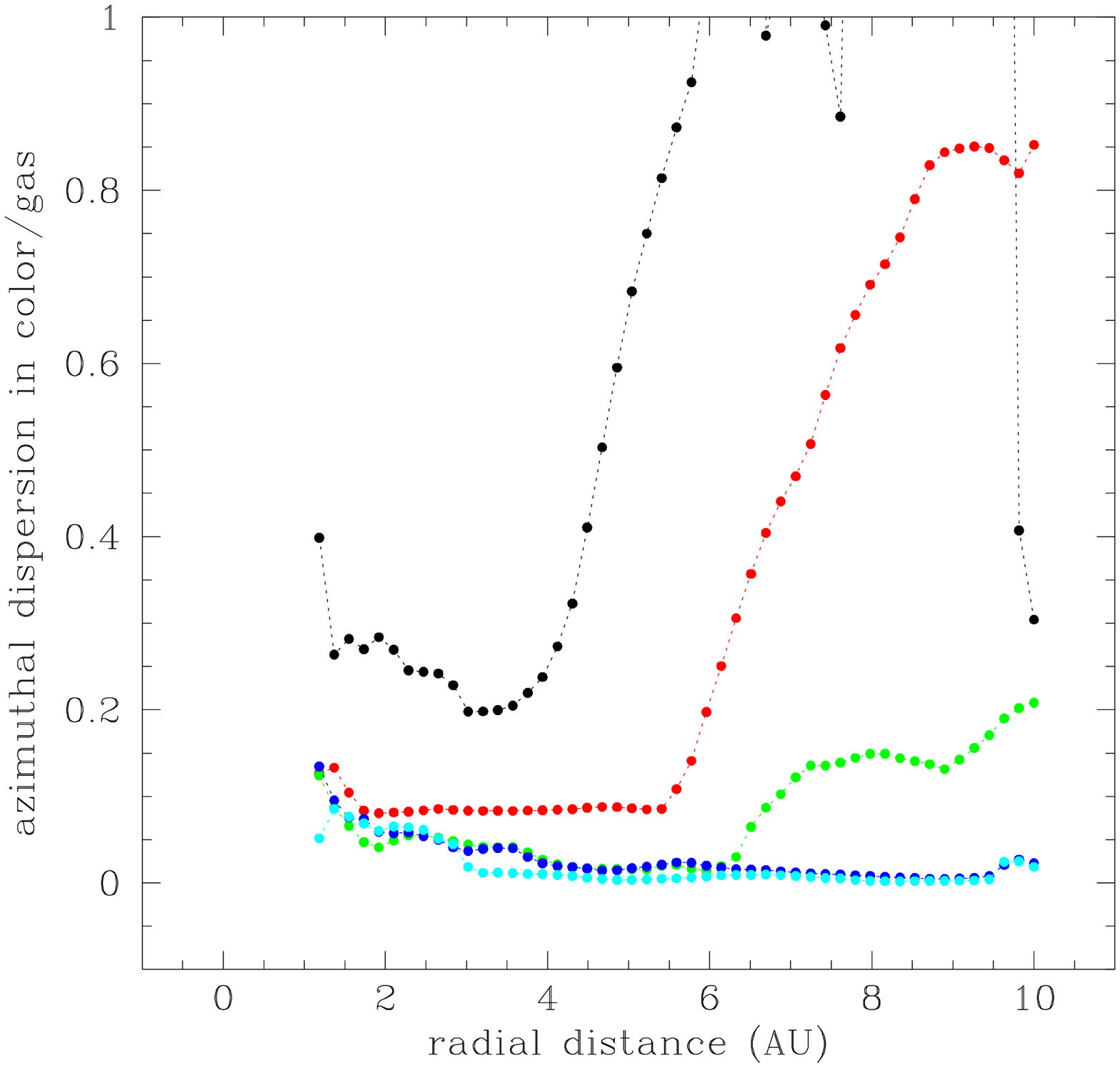}
\caption{Time evolution of the azimuthal dispersion of the color to gas ratio as a 
function of disk radius for model 0.1-1.8-2, with single-shot injection at 2 AU
and a 0.1 $M_\odot$ protostar, at times: black - 71 yr, red - 320 yr, green - 430 yr, 
blue - 5300 yr, and cyan - 8040 yr.}
\end{figure}

\clearpage

\begin{figure}
\vspace{-2.0in}
\plotone{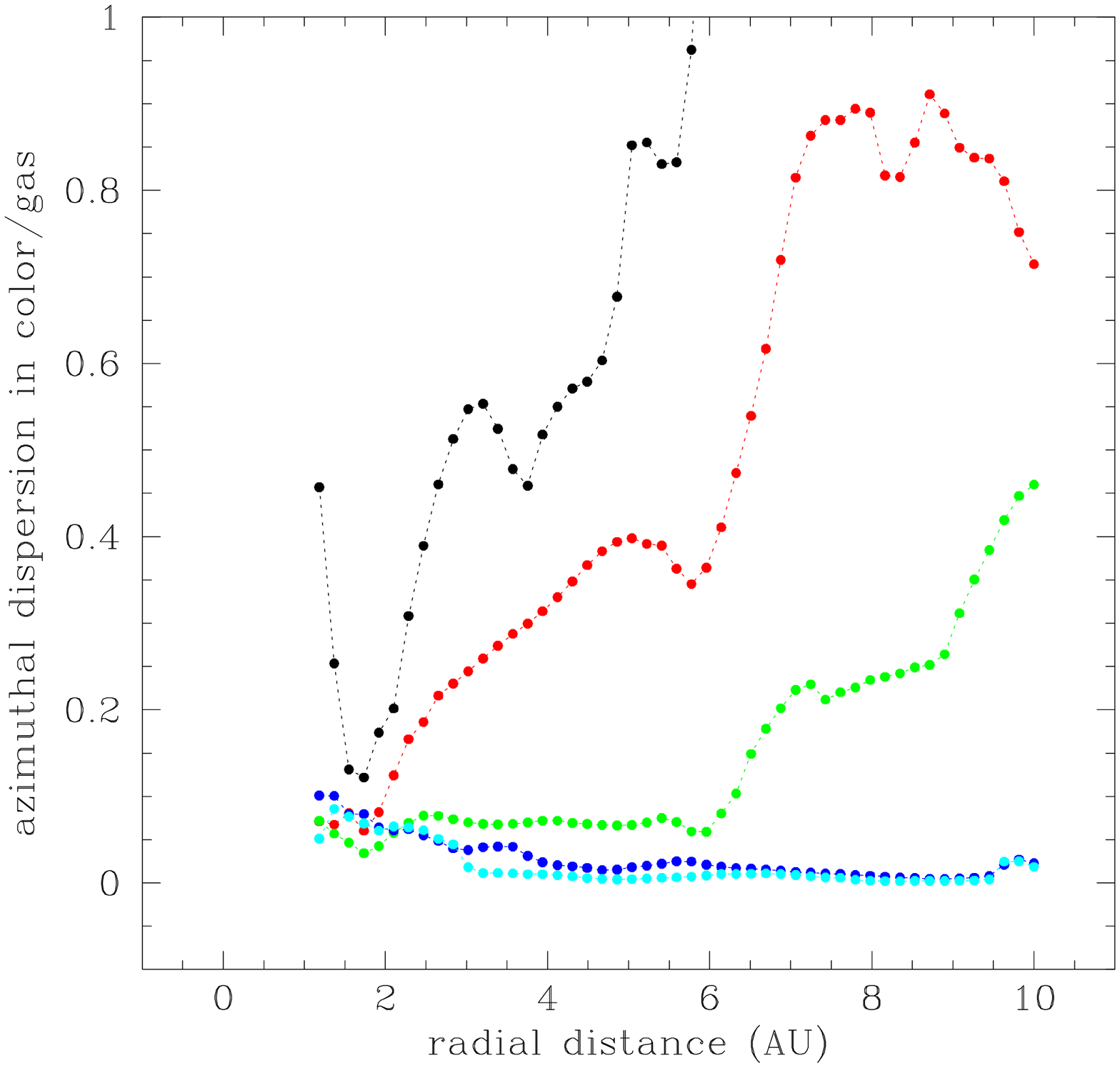}
\caption{Time evolution of the azimuthal dispersion of the color to gas ratio as a 
function of disk radius for model 0.1-1.8-9, with single-shot injection at 9 AU
and a 0.1 $M_\odot$ protostar, at times: black - 140 yr, red - 320 yr, green - 430 yr, 
blue - 5300 yr, and cyan - 8040 yr.}
\end{figure}

\clearpage

\begin{figure}
\vspace{-2.0in}
\plotone{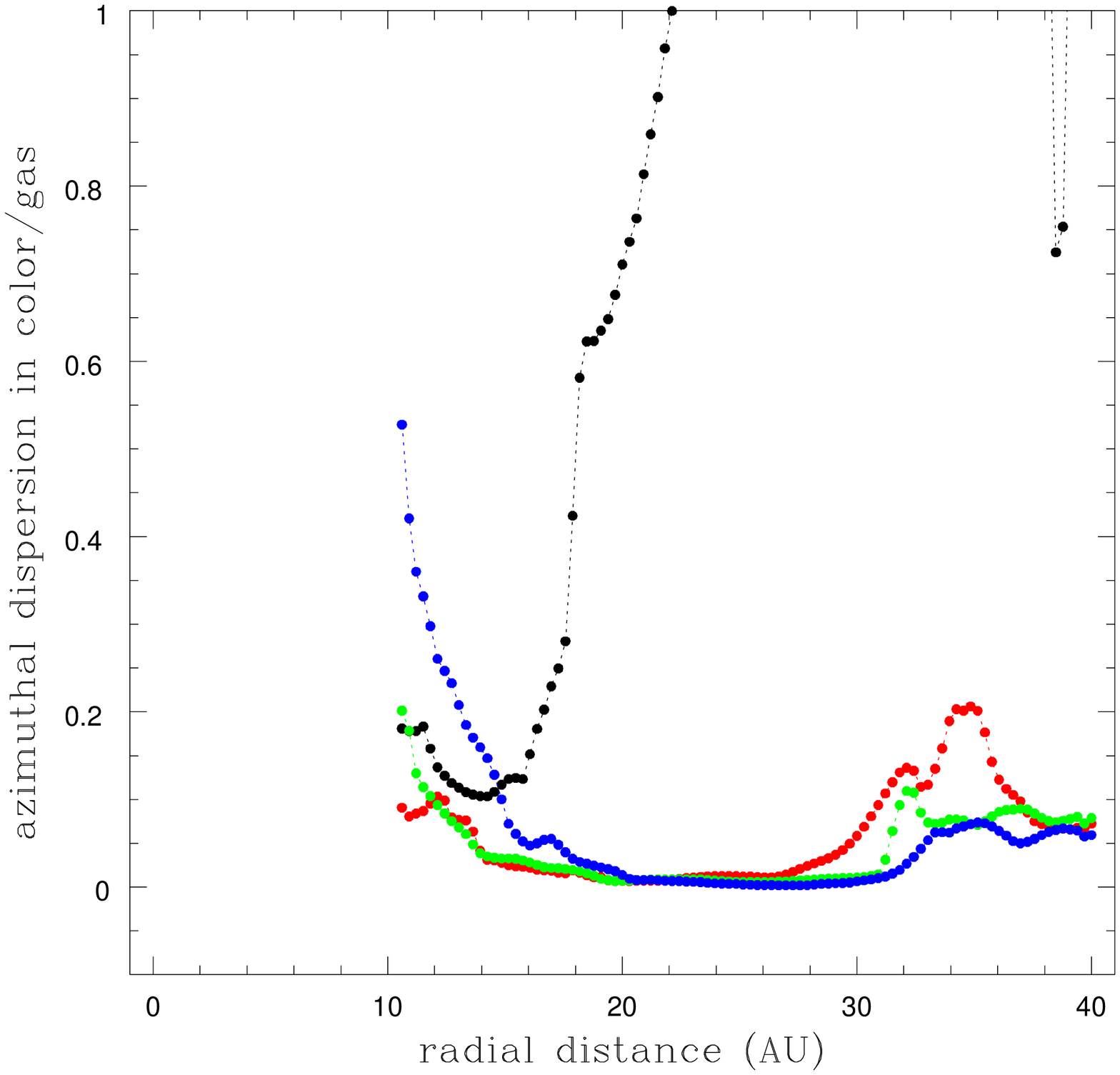}
\caption{Time evolution of the azimuthal dispersion of the color to gas ratio as a 
function of disk radius for model 1.0-1.1-40-20, with single-shot injection at 20 AU
on the surface of a 40 AU radius disk around a 1.0 $M_\odot$ protostar, at times: 
black - 200 yr, red - 5200 yr, green - 12400 yr, and blue - 19700 yr.}
\end{figure}

\clearpage

\begin{figure}
\vspace{-2.0in}
\plotone{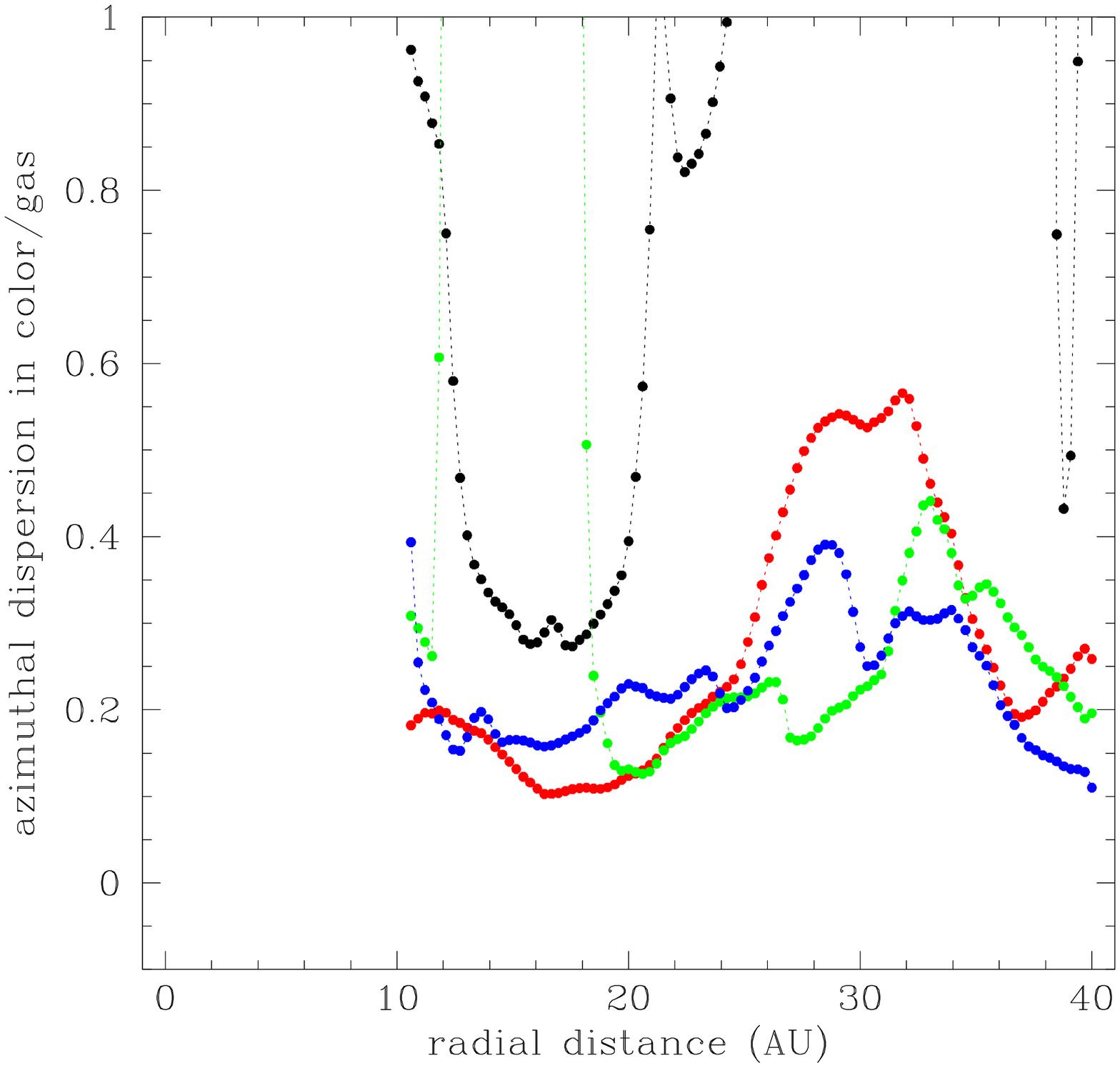}
\caption{Time evolution of the azimuthal dispersion of the color to gas ratio as a 
function of disk radius for model 1.0-1.1-40-20c, with continuous injection at 20 AU
on the surface of a 40 AU radius disk around a 1.0 $M_\odot$ protostar, at times: 
black - 200 yr, red - 5200 yr, green - 12600 yr, and blue - 19600 yr.}
\end{figure}

\clearpage
\begin{figure}
\vspace{-1.0in}
\plotone{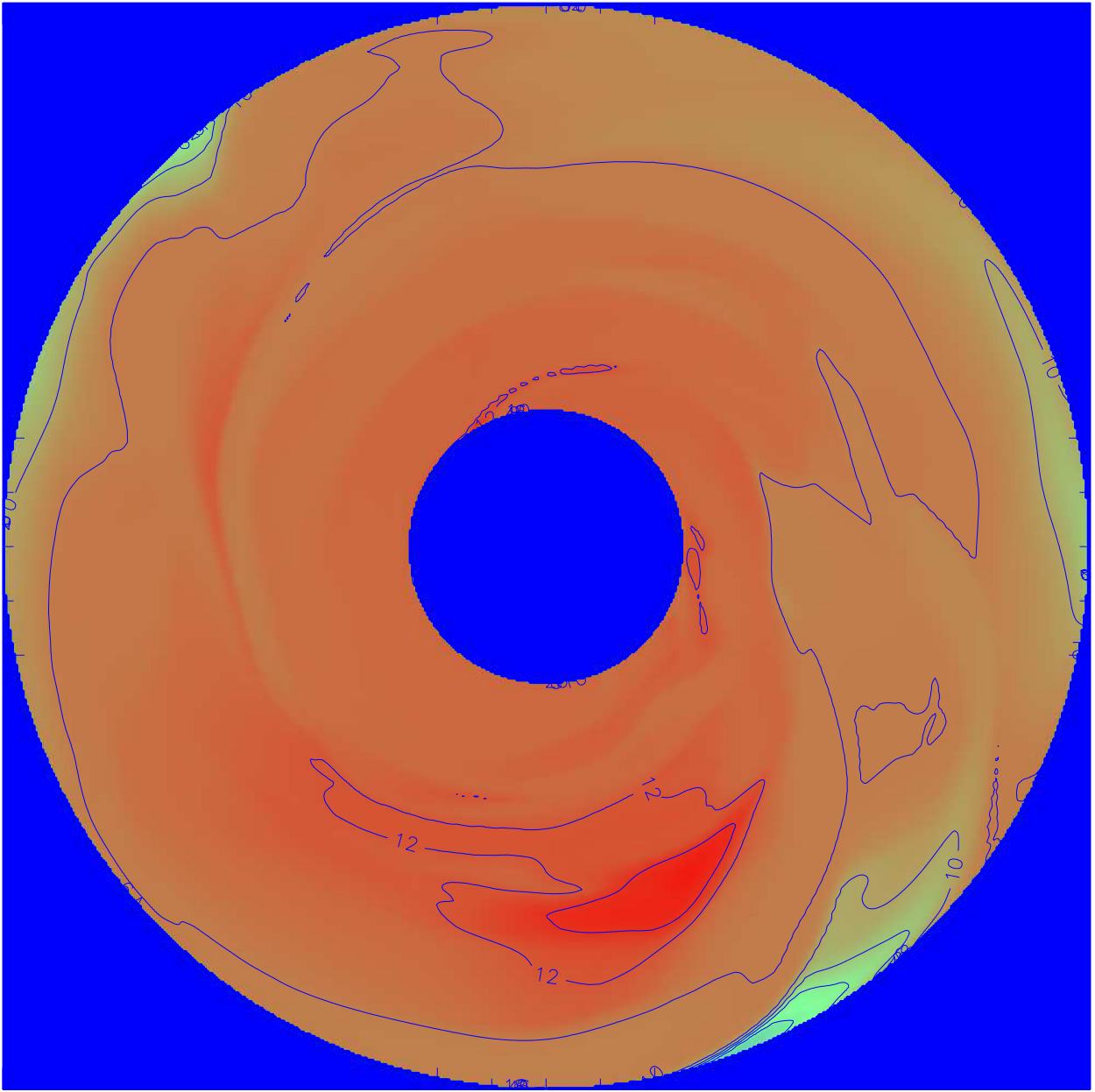}
\caption{Log $\rho_c/\rho$ in the midplane after 405 yr for model 1.0-1.1-40-20c, with 
continuous color injection at the disk's surface at 20 AU. Region
shown is 80 AU in diameter. A 1.0 $M_\odot$ protostar lies at the center of the MGU
disk.}
\end{figure}

\clearpage

\begin{figure}
\vspace{-1.0in}
\plotone{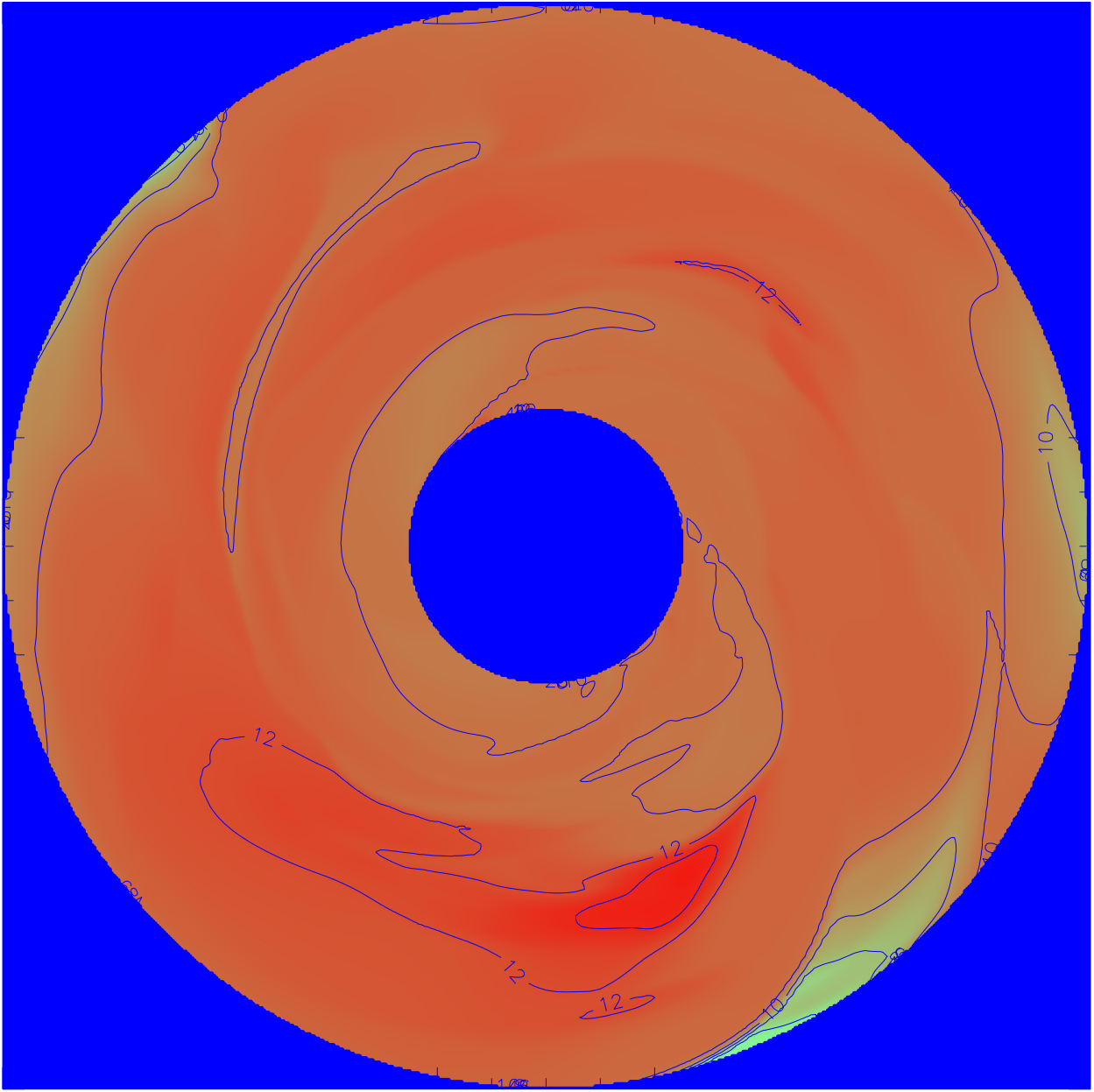}
\caption{Same as Figure 9, except at 405 yr for model 1.0-1.1-40-30c, with 
continuous color injection at the disk's surface at 30 AU, instead of 20 AU.}
\end{figure}

\clearpage

\end{document}